\UseRawInputEncoding
\documentclass[twocolumn,tighten]{aastex62}
\usepackage{natbib}

\usepackage[hang,raggedright,tight]{subfigure}
\usepackage{longtable}
\usepackage[figuresright]{rotating}
\usepackage{float}
\usepackage[toc,page]{appendix}

\usepackage{gensymb}

\usepackage{diagbox}
\usepackage{pbox}
\usepackage{multirow}  

\def\gsim{\;\rlap{\lower 2.5pt
\hbox{$\sim$}}\raise 1.5pt\hbox{$>$}\;}
\def\lsim{\;\rlap{\lower 2.5pt
\hbox{$\sim$}}\raise 1.5pt\hbox{$<$}\;}

\usepackage{xcolor}

\newcommand{\ud}{\mathrm{d}}
\newcommand{\be}{\begin{equation}\rm}
\newcommand{\ee}{\end{equation}}
\usepackage{amsmath}

\shorttitle{The specific merger rate}
\shortauthors{Dong et al.}

\begin{document}
\title{The Universal Specific Merger Rate of Dark Matter Halos}


\author[0000-0003-0296-0841]{Fuyu Dong}
\altaffiliation{dongfy2020@kias.re.kr}
\affiliation{Key Laboratory for Research in Galaxies and Cosmology, Shanghai Astronomical Observatory, Shanghai 200030, China} 
\affiliation{School of Physics, Korea Institute for Advanced Study (KIAS), 85 Hoegiro, Dongdaemun-gu, Seoul, 02455, Republic of Korea}
\affiliation{Department of Astronomy, School of Physics and Astronomy, Shanghai Jiao Tong University, Shanghai, 200240,China}

\author{Donghai Zhao}
\altaffiliation{dhzhao@shao.ac.cn}
\affiliation{Key Laboratory for Research in Galaxies and Cosmology, Shanghai Astronomical Observatory, Shanghai 200030, China} 

\author[0000-0002-8010-6715]{Jiaxin Han}
\altaffiliation{jiaxin.han@sjtu.edu.cn}
\affiliation{Department of Astronomy, School of Physics and Astronomy, Shanghai Jiao Tong University, Shanghai, 200240,China} 
\affiliation{Shanghai Key Laboratory for Particle Physics and Cosmology, Shanghai, 200240, China}
\affiliation{Key Laboratory for Particle Astrophysics and Cosmology (MOE), Shanghai 200240, China}

\author[0000-0001-7890-4964]{Zhaozhou Li}
\affiliation{Key Laboratory for Research in Galaxies and Cosmology, Shanghai Astronomical Observatory, Shanghai 200030, China} 
\affiliation{Department of Astronomy, School of Physics and Astronomy, Shanghai Jiao Tong University, Shanghai, 200240,China}
\affiliation{Shanghai Key Laboratory for Particle Physics and Cosmology, Shanghai, 200240, China} 
\affiliation{Centre for Astrophysics and Planetary Science, Racah Institute of Physics, The Hebrew University, Jerusalem, 91904, Israel}

\author[0000-0002-4534-3125]{Yipeng Jing}
\affiliation{Department of Astronomy, School of Physics and Astronomy, Shanghai Jiao Tong University, Shanghai, 200240,China} 
\affiliation{Shanghai Key Laboratory for Particle Physics and Cosmology, Shanghai, 200240, China} 
\affiliation{Key Laboratory for Particle Astrophysics and Cosmology (MOE), Shanghai 200240, China}
\affiliation{Tsung-Dao Lee Institute, Shanghai Jiao Tong University, Shanghai, 200240, China}

\author[0000-0003-3997-4606]{Xiaohu Yang}
\affiliation{Department of Astronomy, School of Physics and Astronomy, Shanghai Jiao Tong University, Shanghai, 200240,China} 
\affiliation{Shanghai Key Laboratory for Particle Physics and Cosmology, Shanghai, 200240, China} 
\affiliation{Key Laboratory for Particle Astrophysics and Cosmology (MOE), Shanghai 200240, China}
\affiliation{Tsung-Dao Lee Institute, Shanghai Jiao Tong University, Shanghai, 200240, China}


\begin{abstract}
We employ a set of high resolution N-body simulations to study the merger rate of dark matter halos. We define a specific merger rate by normalizing the average number of mergers per halo with the logarithmic mass growth change of the hosts at the time of accretion. Based on the simulation results, we find that this specific merger rate, $\mathrm{d}N_{\mathrm{merge}}(\xi|M,z)/\mathrm{d}\xi/\mathrm{d}\log M(z)$, has a universal form, which is only a function of the mass ratio of merging halo pairs, $\xi$, and does not depend on the host halo mass, $M$, or redshift, $z$, over a wide range of masses ($10^{12}\lesssim M \lesssim10^{14}\,M_\odot/h$) and merger ratios ($\xi\ge 1e-2$). We further test with simulations of different $\Omega_m$ and $\sigma_8$, and get the same specific merger rate. The universality of the specific merger rate shows that halos in the universe are built up self-similarly, with a universal composition in the mass contributions and an absolute merger rate that grows in proportion to the halo mass growth. As a result, the absolute merger rate relates with redshift and cosmology only through the halo mass variable, whose evolution can be readily obtained from the universal mass accretion history (MAH) model of \cite{2009ApJ...707..354Z}. Lastly, we show that this universal specific merger rate immediately predicts an universal un-evolved subhalo mass function that is independent on the redshift, MAH or the final halo mass, and vice versa.
\end{abstract}


\section{INTRODUCTION}
\label{introduction}
Subhalo accretion is one of the key ingredients of the hierarchical structure formation theory. In the $\Lambda$ Cold Dark Matter ($\Lambda$CDM) framework, dark matter halos form from the small density perturbations in the early universe, and then grow in size by accreting surrounding smaller halos under gravity. These mergers are not only responsible for the growth of halos, but also lead to the formation of subhalos within halos. As galaxies form and evolve in the centers of halos, the merger history of a halo also determines the property and distribution of its galaxy population. It is thus important to investigate the composition and rate of halo mergers across cosmic time in order to better understand structure formation and galaxy evolution\citep[e.g.,][]{1993MNRAS.262..627L,2015MNRAS.449...49R}. 

During the past, great efforts have been made to characterize halo merger in a CDM cosmology, either through (semi-) analytical models\citep{1991ApJ...379..440B,1993MNRAS.262..627L,10.1046/j.1365-8711.1999.02477.x,2001MNRAS.323....1S,2002MNRAS.329...61S,2008MNRAS.389.1521Z,2008MNRAS.388.1792N,2011ApJ...741...13Y,2021ApJ...914..141S,SS21}
or numerical simulations. Due to the complexity of the structure formation process, the latter allows for a more accurate and detailed study \citep{1999AJ....117.1651G,2001ApJ...546..223G,2006ApJ...652...56B}, especially in recent years with the rapid advances in computer technology. For example, \cite[FM08 afterwards]{2008MNRAS.386..577F} has proposed a fitting formula of the mean merger rate per halo $\mathrm{d}N_{\mathrm{merge}}/N_{\mathrm{halo}}/\mathrm{d}z/\mathrm{d}\xi$ based on FoF halos in the Millennium Simulation, where $z$ is redshift, $\xi$ the mass ratio of merging halo pairs. Their follow-up works can be found in \cite{2009MNRAS.394.1825F} and \cite[here after FM10]{2010MNRAS.406.2267F}. They found that the mean merger rate per halo has low dependence on the halo mass $M^{0.13}$ and redshift $(1+z)^{0.099}$.

Although simulation is more reliable in capturing the detailed process of gravitational collapse in principle, the measurement of merger rate could be subject to a lot of tricky issues such as the construction of the merger tree\citep[e.g.][]{Srisawat13,2018MNRAS.474..604H} and the way to identify halos and subhalos\citep[e.g.][]{Muldrew11,2012MNRAS.427.2437H, Onions12, Knebe13}.
As a result, some discrepancies still remain among the simulation measurements of the merger rates (e.g.,\citealp[hereafter G09]{2009ApJ...701.2002G}; \citealp[hereafter G10]{2010ApJ...719..229G}; \citealp{2009MNRAS.395.1376W}; \citealp[hereafter S09]{2009ApJ...702.1005S}; \citealp{2017MNRAS.472.3659P}). Among these studies, G09 and S09 have found a more
rapid increase in the merger rate with redshift  at $z<1$ compared to FM08 and FM10,
while the discrepancies  at higher redshift become reasonably small. Consequently, the fitting formulas of merger rate proposed in different works still lead to differing dependencies upon the halo mass, redshift, and cosmological parameters.

In this work, we propose a model that describes the halo merger rate in a simpler way without involving parameters related with the halo itself or cosmology. This is done by
normalizing the merger rate per halo with the logarithmic-mass growth rate of the host halo:
\begin{equation}
\mathrm{d}N_{\mathrm{merge}}(\xi|z,M)/\mathrm{d}\xi/\mathrm{d}\log M=\frac{\mathrm{d}N_{\mathrm{merge}}/\mathrm{d}z/\mathrm{d}\xi}{\mathrm{d}\log M/\mathrm{d}z},
\end{equation}
where $\mathrm{d}\log M$ represents the mass growth ratio of the host, $\log ((M+dM)/M)\sim dM/M$, between redshift $z$ and $z+\mathrm{d}z$. 
It describes the transient populations of infall subhalos relative to the host halo.
We will show that this specific merger rate has a universal form for halos of different masses and redshifts and is only a function of the merger mass ratio $\xi$. Moreover, we look at simulations with different $\Omega_m$ and $\sigma_8$, and find the form of the specific merger rate remains the same.


In addition to the above instantaneous merger rate at different redshifts, there are also amount of studies discussing the merger histories of present day halos. For example,
\cite{2005MNRAS.359.1029V} has found that the accumulated un-evolved subhalo mass function is approximately universal for halos with different masses, which has been later confirmed by other studies \citep{2008MNRAS.386.2135G,2008ApJ...683..597S,2011ApJ...741...13Y,2014MNRAS.440..193J,2016MNRAS.458.2848J,2018MNRAS.474..604H}.
We will show that we can naturally derive this universality from our specific merger rate. 

Contrary to a common consensus that halo growth can be divided into an early and a late phase that are contributed by statistically different types of mergers, our results suggest that these two phases are actually composed of the same types of mergers. As the density profile of a halo is shaped by its growth history, this means the types of merger should not be responsible for the different slopes of the inner and outer profiles of the halo. This is consistent with an early claim that the density profile of a halo is not determined by the types of mergers it experienced (i.e., aggregation history~\citealp{2005MNRAS.358..901S}).


In this work, we use a set of dark matter only simulations to study the halo merger rate. Based on the simulation results, we show that the Mass Accretion Histories (MAHs) of halos of different masses and cosmologies are composed of mergers of 
the same
distribution in mass ratio. 
Compared to previous studies, our finding is simpler in the description and closer to the physics behind.
This paper is organized as follows: In \S\ref{sec:data}, we describe the simulation data and introduce the specific merger rate. Our main results are given in \S\ref{sec:main}.  We discuss the connections of our result to other merger statistics in \S\ref{sec:apply} and compare with theoretical expectations from the excursion set model in \S\ref{sec:EPS}. We conclude in \S\ref{sec:summay}. Throughout this work, we use $\log $ for 10-based logarithm and $\ln$ for natural logarithm.

\section{Data and Methodology}
\label{sec:data}
\subsection{Simulation}

\begin{table*}
\centering
\caption{Simulation Parameters.
  \label{Table1}}
\begin{minipage}{180mm}
\centering
\begin{tabular}{lllllllll}
\hline
{Simulation} & {box size} &{particle mass}& $\Omega_\Lambda$ & $\Omega_m$ & $\Omega_b$ & h &$\sigma_8$ &$n_s$   \\
 & ($h^{-1} Mpc$) & ($h^{-1} M_\odot$) &  &  &  &  &  &  \\
\hline
{L1} &  150 & 2.34$\times10^8$ &  0.732 & 0.268 & 0.045 & 0.71 & 0.85 & 1 \\
{L2} &  300 & 1.87$\times10^9$ &  0.732 & 0.268 & 0.045 & 0.71 & 0.85 & 1\\
{L3} &  100 & 6.67$\times10^7$ &  0.742 & 0.258 & 0.044 & 0.719 & 0.796 & 0.963\\
{L4} &  300 & 1.87$\times10^9$ &  0.732 & 0.268 & 0.045 & 0.71 & 0.95 & 1\\
{L5} &  300 & 1.39$\times10^9$ &  0.8 & 0.2 & 0.045 & 0.71 & 0.85 & 1\\
{EDS}  &  100 & 2.59$\times10^8$ & 0.  & 1.   &0.  &-- &0.796 & 0.963\\
\hline
\end{tabular}
\end{minipage}
\end{table*}

Our work is based on a set of high resolution $\Lambda$CDM N-body simulations \citep{2007ApJ...657..664J}(Table.\ref{Table1}), which are run with the same number of particles $N_P=1024^3$. We respectively label them as Lx (x=1,2,3,4,5, L for $\Lambda$CDM). Among these simulations L1 and L2 have the same cosmology but different box sizes: 150 Mpc/h and 300 Mpc/h.  L3 has the highest mass resolution and lower $\Omega_m$ and $\sigma_8$ compared to L1. L4 has the same mass resolution with L2 but a higher $\sigma_8$ (0.95), and L5 has the same box size with L2 but smaller $\Omega_m$. We also consider an Einstein–de Sitter (EDS) simulation for which $\Omega_m$= 1 and $\sigma_8$ and the shape of power spectrum are the same as L3. These simulations will allow us to study the possible cosmology dependence and analyze resolution issues.
For each simulation, there are $\sim$100 snapshot outputs between redshift 17 and zero.

In all simulations, halos are identified using the standard Friends-of-Friends \citep{1985ApJ...292..371D} algorithm with a linking length equal to 0.2 times the mean particle separation. Based on the FoF halos, subhalos are then identified with the Hierarchical Bound-Tracing \citep{2012MNRAS.427.2437H} algorithm. 
The mass of a self-bound subhalo in HBT is defined as its number of bound particles multiplied by the particle mass.

\subsection{Merging Halo Pairs}
The virial mass of a halo is defined at the radius where its inner matter density equals to the critical density predicted from the spherical collapse model, $\rho_{\rm vir}=\Delta_{\mathrm{vir}}\rho_{\mathrm{crit}}$, where $\Delta_{\mathrm{vir}}=18\pi^2+82[\Omega_m(z)-1]-39[\Omega_m(z)-1]^2$ according to \citet{1998ApJ...495...80B}.
If a satellite subhalo is located within the virial radius of its host halo at any given moment but is out of the virial radius at its previous snapshot, we define it as an ``infall" event. On the contrary, if a satellite subhalo is in the virial radius at the previous snapshot but outside the virial radius at the given moment, we regard it as a ``splashout" event. For an ``infall" event, we define $\xi$ as the mass ratio of the pair of merging halo progenitors. And for a ``splashout" event, $\xi$ is defined as the mass ratio of the halo pair right after the splashout. 

We use the virial masses of the halo pair when calculating the merger mass ratio.
However, it can be non-trivial to define virial masses for halos that are merging. Even though the centers of the progenitors are not contained in the virial radius of each other, their boundaries can still be overlapping.
At this point, we have defined two types of halos: ``independent halos" and ``intersecting halos". A ``independent halo" does not intersect with any other larger halo in virial radius, while the ``intersecting halo" partially overlaps with at least one other larger halo in space. In the latter case, we calculate the virial mass of the larger halo in the normal way according to the spherical overdensity definition, while the virial mass of the smaller halo is computed excluding particles within the virial radius of the larger halo, similar to the treatment of \cite{2008MNRAS.386.2135G}. The redefined halos are always located at the centers of HBT identified self-bound subhalos.
In this work, we only consider the independent halo sample when selecting merged halos for analysis of the merger rate. 
Clearly, the mass of the accreted subhalo will be underestimated in our treatment if it is a intersecting halo in the last snapshot. In this case, we use the viral mass of its progenitor in the last-last snapshot before merger\footnote{We find that the progenitor mass ratio measured at the last-last snapshot is already converged.} to get the mass ratio $\xi$. Besides, we only consider the merging between the first-order subhalos labeled by HBT and the host halo and have not accounted for mergers between higher-level subhalos and the host, as higher level mergers can be analytically modelled subsequently once first-order merger rates are modelled.

\subsection{The Specific Merger Rate}
\label{subsec:normnm}
During the merger, the infalling halos might fall into and run out of the virial radius of host halos more than once. So for a halo with mass $M_h$ at redshift $z_1$, we define its number of ``merger" events between two neighboring snapshots ($z_1$ and $z_2$) as the number of ``infall" events subtracting the ``splashout" events: $\Delta N_{\mathrm{merge}}(\xi|M_h)=\Delta N_{\mathrm{infall}}-\Delta N_{\mathrm{splash}}$.
Here the ``splashout" has an opposite definition of ``infall", referring to the events that the subhalo flies out of the main halo.
Meanwhile, we define the mass change ratio of the main branch halo as: $\Delta \log M_h=\log (M_h(z_1)/M_h(z_2))$, where $z_1<z_2$.  By averaging the results of all the hosts with masses $M_h$, we then statistically obtain the ``specific merger rate" from the simulation: $\langle \Delta N_{\mathrm{merge}}(\xi|M_h,z)\rangle/\langle \Delta \log M_h\rangle / \Delta\xi$. By further integrating this expression on $\xi$ over the range of $\xi>\xi_{\mathrm{min}}$, we could obtain the cumulative form: $\langle \Delta N_{\mathrm{merge}}(>\xi_{\mathrm{min}}|M_h,z)\rangle/\langle \Delta \log M_h\rangle$.

The ``specific merger rate" is the main quantity we study in this paper. In the following, we will use the simulation data to show that the (cumulative) specific merger rate is universal for halos with different masses, redshifts, and cosmologies. 

\section{Result}
In this section, we mainly study the instantaneous specific merger rate for halos of different masses at $z\ge0$ using $\Lambda$CDM simulations. Besides, we also show that studying the merger rate along the evolution history of a specific population of halos produces the same result.

\label{sec:main}
\subsection{The Universality of the Specific Merger Rate}
\label{sec:nm-z0}
\begin{figure*}
    \centering
	  \subfigure{
     \includegraphics[width=0.48\linewidth, clip]{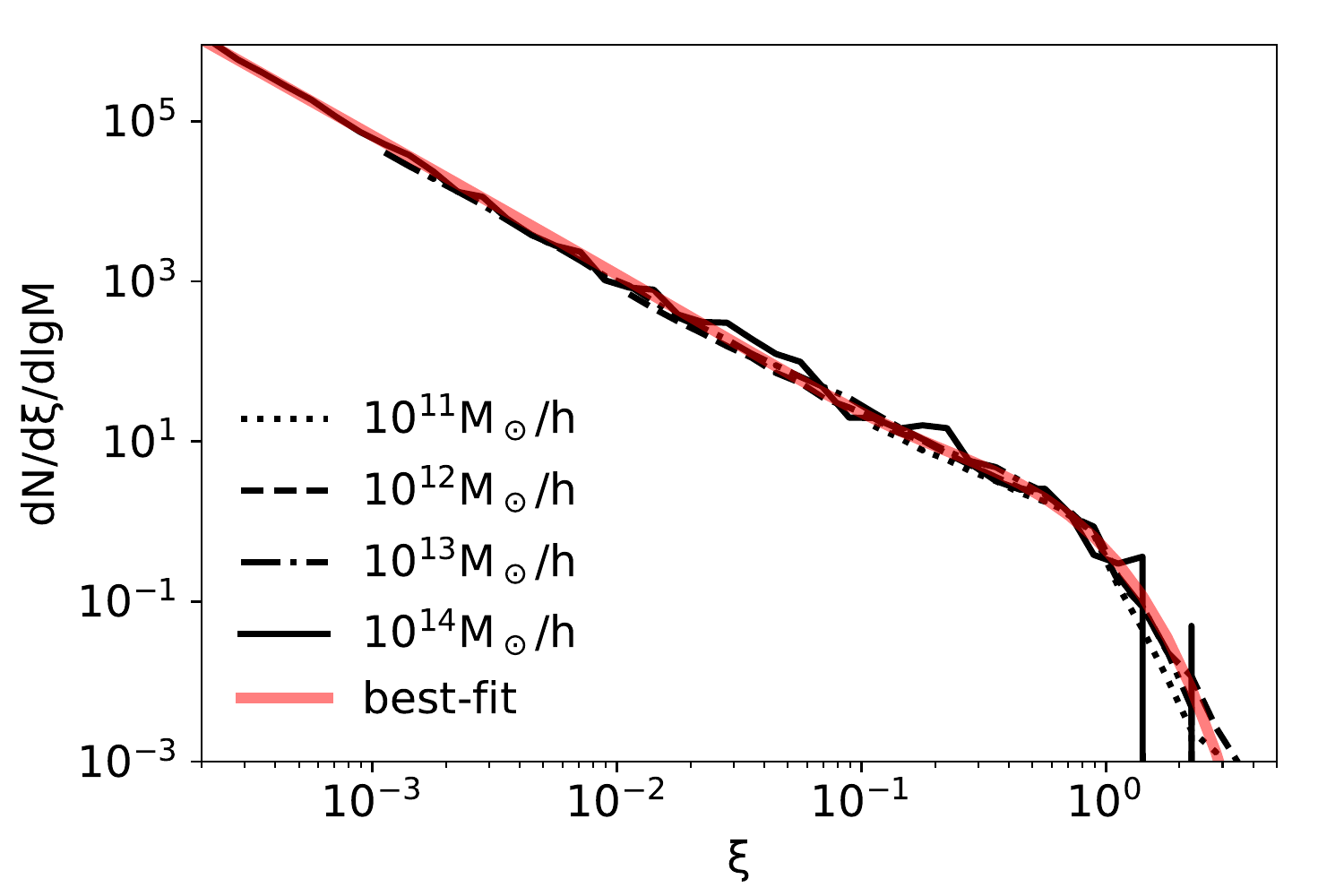}}
     \subfigure{
     \includegraphics[width=0.48\linewidth, clip]{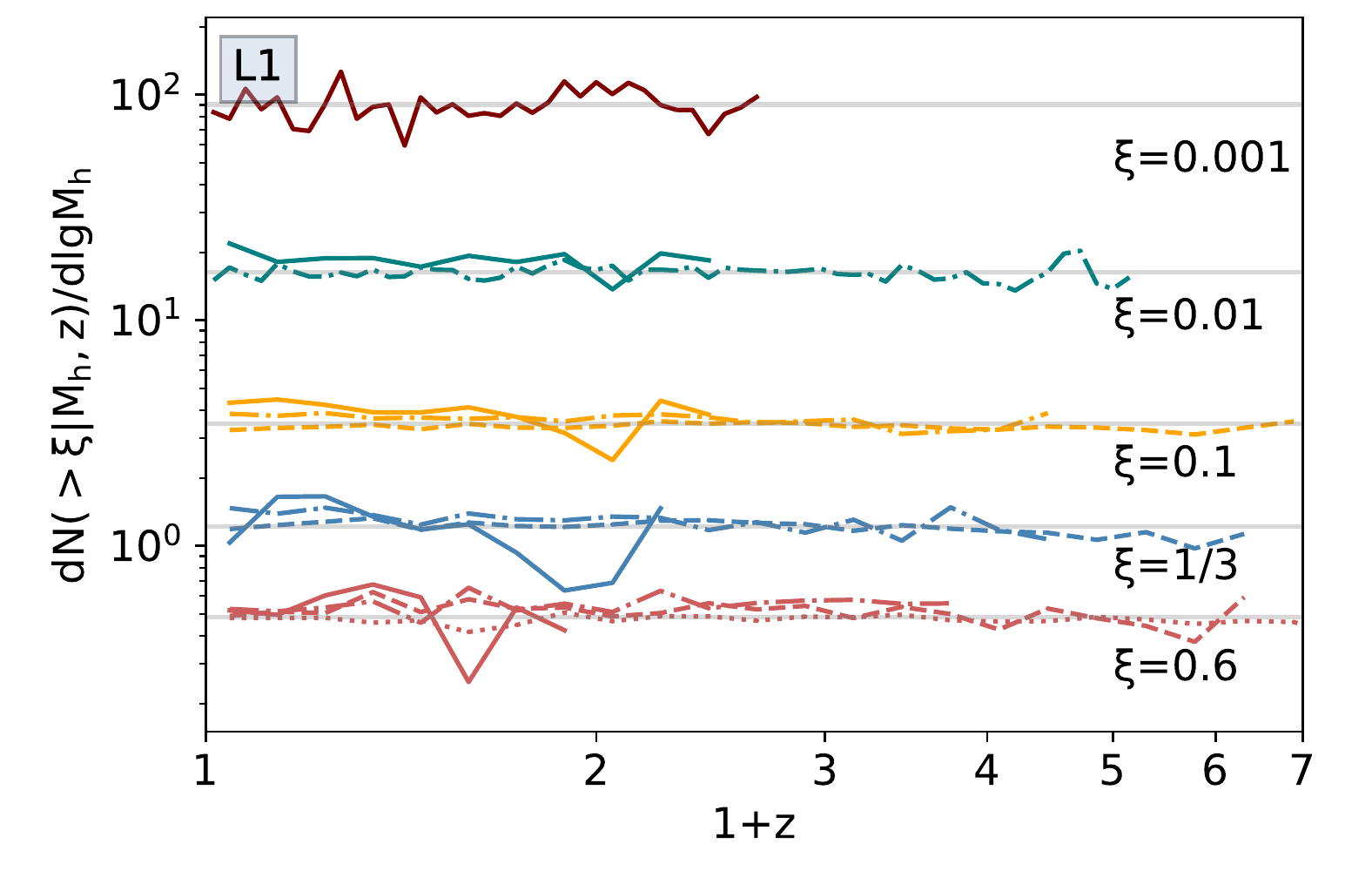}}
\caption{The specific merger rate measured for simultion L1. Left panel: the specific merger rate as a function of mass ratio $\xi$ for halos with different masses at z=0. The red solid line shows the double Schechter  fitting curve of the simulation data. Right panel: the cumulative specific merger rate as a function redshift. Results for different masses are shown in different line-styles. There are five choices of the lower limit $\xi_{\rm min}$ of integration: 0.6, 1/3, 0.1, 0.01 and 0.001, for which the measurements are presented in different colors. The grey solid lines in the right panel are obtained by integrating the double Schechter fitting curve down to different values of $\xi_{\rm min}$.}
    \label{fig:NM-150}
\end{figure*}

\begin{figure*}
    \centering
     \subfigure{
     \includegraphics[width=1\linewidth, clip]{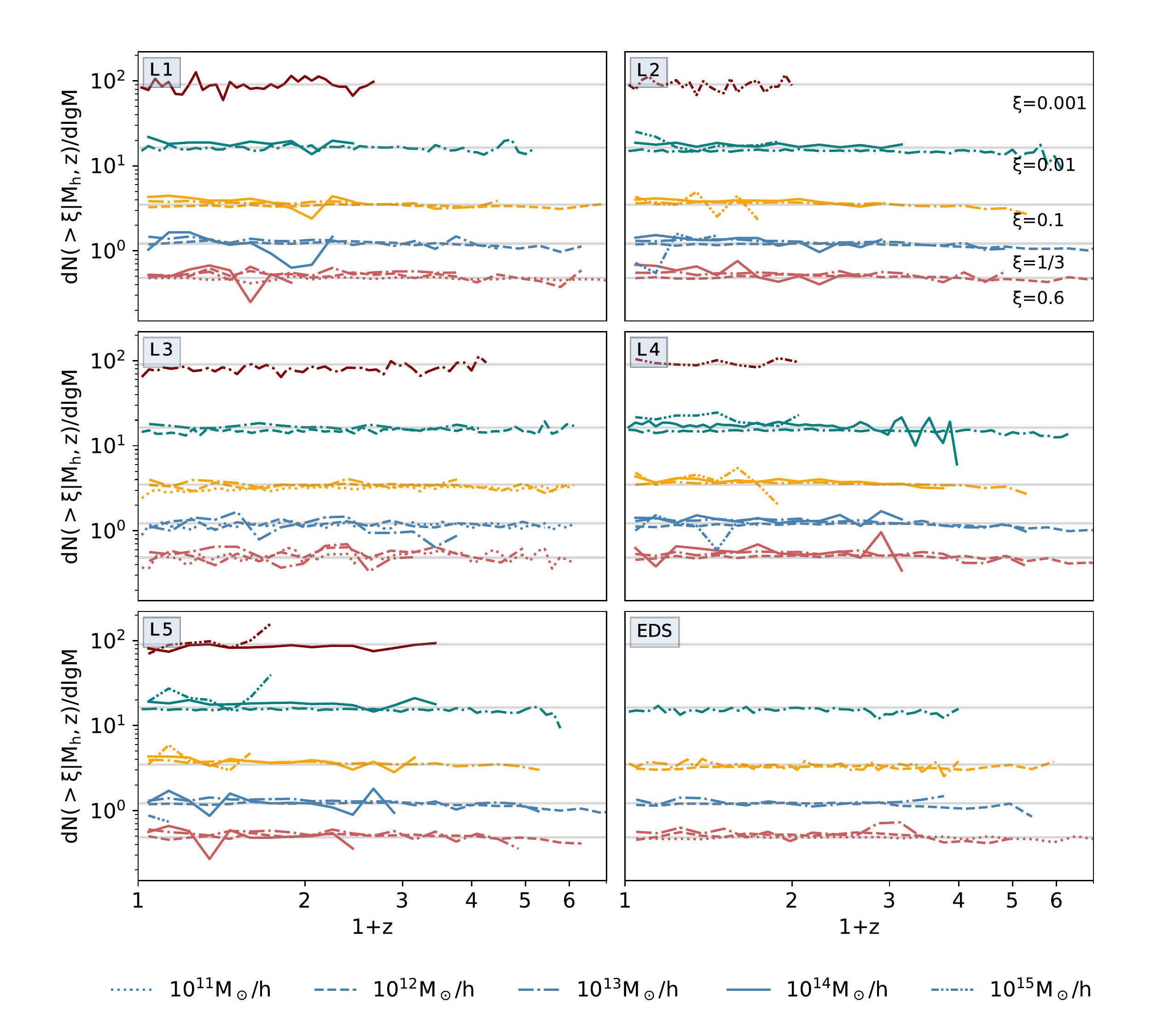}}
   \caption{The cumulative specific merger rates measured for different simulations. The six panels from left to right and top to bottom are respectively for simulation L1, L2, L3, L4, L5 and EDS. The line-styles and colors in this figure are the same as in Fig.\ref{fig:NM-150}.}
    \label{fig:NM-123456}
\end{figure*}


In simulation, we bin halos at each snapshot into four mass ranges for statistics: $10^{11}M_\odot/h$, $10^{12}M_\odot/h$, $10^{13}M_\odot/h$ and $10^{14}M_\odot/h$.
For each halo at a given snapshot, we find out all the new mergers occurred after the last snapshot. Meanwhile, we calculate the average mass change ratio $\Delta \log M_h$ of halos of the same mass. And the instantaneous specific merger rate is obtained by normalizing the average number of mergers with the mean mass change ratio $\Delta \log M_h$.
In Fig.\ref{fig:NM-150} we show one of such measurements with simulation L1.  The left panel shows the results at redshft zero. From the figure we see that the specific merger rates for different mass halos are well consistent with each other, and a double Schechter \citep{2018MNRAS.474..604H} provides a good description:
\begin{equation}
\label{eq:DS-fit}
\mathrm{d}N_{\rm merge}/\mathrm{d}\xi/\mathrm{d}\log M_h=(a_1\xi^{b_1}+a_2\xi^{b_2})\exp(c\xi^d),
\end{equation}
where $a_1$, $a_2$, $b_1$, $b_2$, c and d are free parameters. The fitting curve is shown in the red solid line with the best-fit parameters to be $(a_1,a_2,b_1,b_2,c,d) = (0.467, 17.362, -1.717, 0.212, -3.682, 0.937)$.
For the numerical results in the figure, we make a lower cut-off in the mass ratio considering the mass resolution effect.

Next, we measure the specific merger rates at different redshift. In order to show the redshift dependence more clearly, we use the cumulative form of the specific merger rate in the following of the paper, which is obtained by integrating $\mathrm{d}N_{\mathrm{merge}}/\mathrm{d}\xi/\mathrm{d}\log M_h$  over $\xi>\xi_{\mathrm{min}}$. Here we make five choices of $\xi_{\mathrm{min}}$ of integration for analysis: 0.6, 1/3, 0.1, 0.01 and 0.001. The results are shown in the right panel of Fig.\ref{fig:NM-150}, in which the grey solid lines are achieved by directly integrating the fitting formula of EQ.\ref{eq:DS-fit}. 
The colored lines are simulation results, which are smoothed by averaging both $\langle\Delta N_{\mathrm{merge}}(\xi>\xi_{\mathrm{min}},z)\rangle$ and $\langle\Delta \log M(z)\rangle$ every three adjacent data points whenever the fluctuation of the curve exceeds $20\%$ of the mean.
For a given $\xi_{\mathrm{min}}$, we find that the cumulative merger rate is almost a constant over redshift. This discovery is of great significance, implying that the normalized form of the merger rate likely has no dependence on cosmology, as the cosmological parameters vary significantly with redshift. 

As a further illustration, we repeat the above analysis on the other four LCDM simulations which have different $\Omega_m$ and $\sigma_8$ values from L1. Besides, we also consider an EDS simulation ($\Omega_m=1$) for comparision. These results are presented in Fig.\ref{fig:NM-123456}, in which all the simulations show good consistency with each other and with our model, and therefore can be seen as a robust test of the universality of the specific merger rate.

Among these simulations, L3 has the best mass resolution and thus is able to show merger rate down to $\xi=0.01$ for $M=10^{12}M_\odot/h$ halos. With the L3 simulation, we also find that the merger rate for $M_h = 10^{11} M_\odot/h$
and $\xi=0.1$ overlaps with other lines, indicating that our
conclusion might be valid for very low mass halos.
Note that in our cumulative merger rate statistics above, the minimum number of particles of a host halo is about 320. In addition, we only consider pairs in which the less massive progenitor has more than $\sim$30 particles.

Overall, we see general remarkable agreement between our model and simulation results. All the clues above show that the specific merger rate is independent of halo mass, redshift and cosmology at least over the range of halo mass $[10^{12}, 10^{14}]M_\odot/h$, mass ratio $\xi\geq 0.01$ and redshift [0, 5]. In section~\ref{sec:EPS}, we will discuss how this can be approximately understood in the framework of the Extended Press-Schechter (EPS) theory.

\subsection{The Specific Merger Rate along the Evolution History of a Given Halo Population}
\label{sec:history}
\begin{figure}
    \centering
    \subfigure{
     \includegraphics[width=1\linewidth, clip]{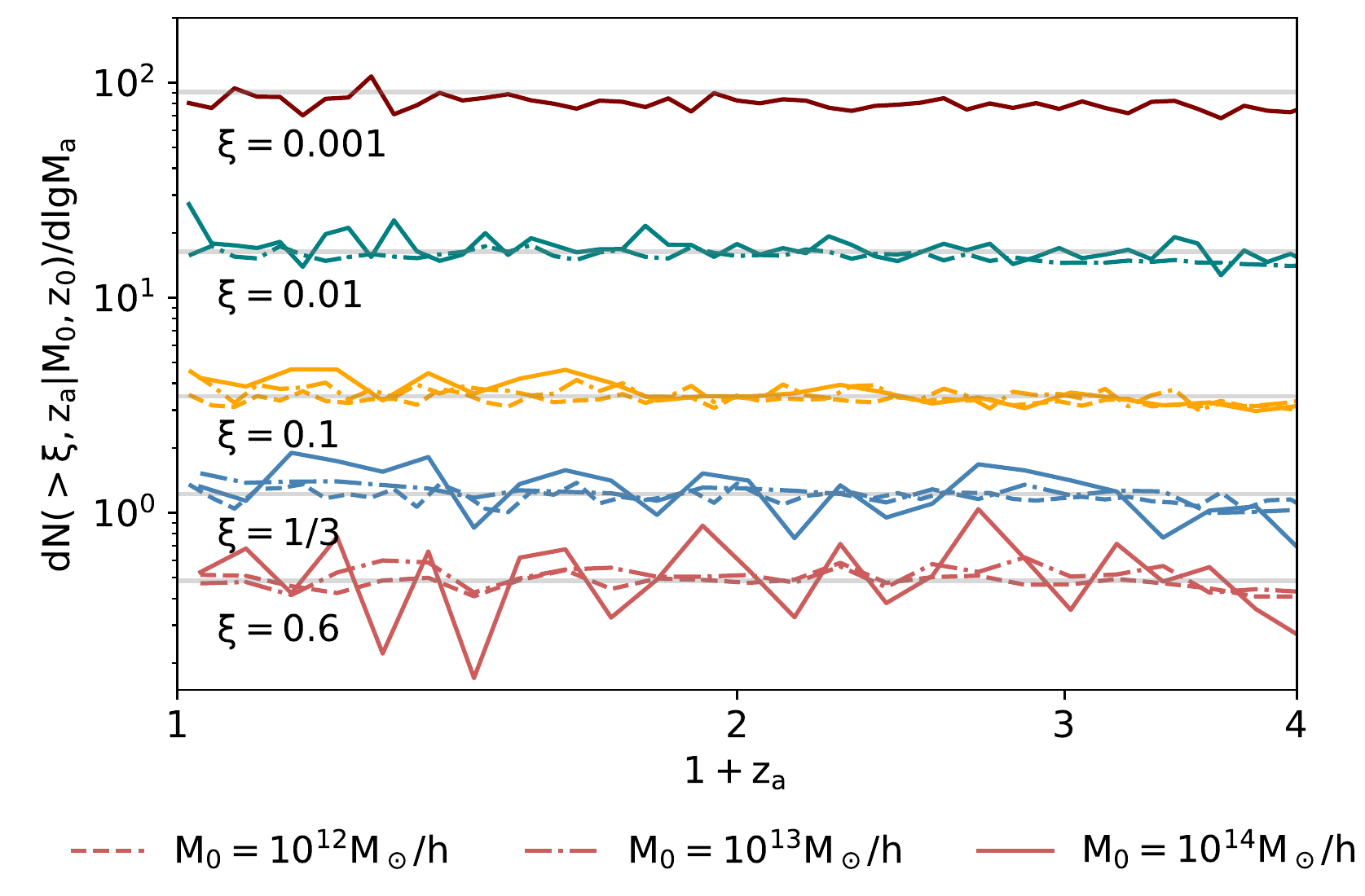}}
   \caption{The cumulative distribution of merger ratios across the histories of present-day halos measured with simulation L1. Results for different present-day masses $M_0$ are shown in different line-styles. The color in this figure has the same meaning as in Fig.\ref{fig:NM-150}.}
    \label{fig:history-nm}
\end{figure}

In above sections, we consider halos of a given mass at different redshifts. Now we trace present-day halos of the same masses along their formation histories. This is particularly interesting for semi-analytical models. 

Assuming $M_0$ is the mass of halos at redshift zero ($z_0$), and $M_a(z_a|M_0,z_0)$ is the main-branch halo mass at redshift $z_a$, then the specific merger rate of $\xi$ at redshift $z_a$ can be obtained by dividing the number of merger events in $[z_a,z_a+dz]$  with the main-branch mass change ratio:
$\langle \Delta N_{\mathrm{merge}}(\xi,z_a|M_0,z_0)\rangle/\langle \Delta \log M_a\rangle$,
where $\langle \Delta \log M_a\rangle=\langle \log [M_a(z_a|M_0,z_0)/M_a(z_a+dz|M_0,z_0)]\rangle$, $\xi=M_s/M_a$ and $M_s$ the mass of accreted subhalo.
The specific merger rate as function of redshift is shown in Fig.\ref{fig:history-nm}. 
As this statistics is more susceptible to the simulation resolution effect compared to the instantaneous merger rate, we only show the results for redshift less than 3.  
It is reasurring to see that we get exactly the same result as in \S\ref{sec:nm-z0}, consistent with universal form of the instantaneous merger rates of halos at any redshift and halo mass.


\section{Connections to other Merger Statistics}
\label{sec:apply}
In this section, we discuss how our model is connected to the un-evolved subhalo mass function and the absolute halo merger rate studied in other works. We also compare the predicted rate of major mergers to other studies in the literature.

\subsection{Equivalence to the Universal Un-evolved Subhalo Mass Function}
An immediate application of the specific merger rate is to predict the un-evolved subhalo mass function (USMF) $g(\xi_0)=\mathrm{d}N(\xi_0)/\mathrm{d}\xi_0$, where $\xi_0$ is the ratio between  the mass of the subhalo at the time of accretion and the present-day host halo mass, such that $\xi_0=\xi(z) M(z)/M_0$. Rewritting the specific merger rate as\footnote{Note the $f(\xi)$ defined here is larger than Equation~\eqref{eq:DS-fit} by a factor of $\ln 10$ due to the different logarithm used.}
\begin{equation}
\begin{aligned}
 f(\xi)&\equiv\frac{\mathrm{d}N(\xi|M,z)}{\mathrm{d}\xi \mathrm{d}\mathrm{ln}M},\\
 &=\frac{\mathrm{d}N(\xi_0/\tilde{M}|\tilde{M},z)}{\mathrm{d}(\xi_0/\tilde{M}) \mathrm{d}\mathrm{ln}\tilde{M}},\\
\end{aligned}
\end{equation}
we can integrate it over the mass increment to obtain the final USMF
\begin{equation}
g(\xi_0)=\int_0^1\frac{f\left(\xi_0/\tilde{M}\right)}{\tilde{M}^2} d\tilde{M}, \label{eq:unevolvedSMF}
\end{equation}
where $\tilde{M}\equiv M(z)/M_0$, with $M(z)$ being the host halo mass at redshift $z$ and $M_0$ the present-day mass. 
Alternatively, one can differentiate the USMF with respect to $M$ to derive the specific merger rate, obtaining 
\begin{equation}
    f(\xi)=\frac{\mathrm{d} G(\xi)}{\mathrm{d}\xi},\label{eq:unevolvedSMF2}
\end{equation} where $G(\xi)\equiv\mathrm{d}N/\mathrm{d}\ln\xi= \xi g(\xi)$. To get the universal specific merger rate, the universality of the USMF over different time is required. Note Equation~\eqref{eq:unevolvedSMF2} is the mathematical inversion of Equation~\eqref{eq:unevolvedSMF}. 

Because the specific merger rate is independent on the host halo mass and redshift, Equation~\eqref{eq:unevolvedSMF} immediately leads to an important conclusion that the USMF is universal, which only depends on the mass ratio $\xi_0$ and is independent on the mass, redshift or MAH of the host halo. This strong universality is a direct consequence of our finding that the halo accretion closely traces the mass accretion, so that the accumulated population of accreted halos does not depend on the path of the mass growth, in line with the ``unbiased accretion" picture discussed in \citet{2016MNRAS.457.1208H}. In fact, multiplying our specific merger rate by $\xi$, one get 
\begin{equation}
    \frac{m\ud N}{\ud\xi\ud M}=\xi f(\xi),
\end{equation} which means the mass accreted through mergers at a fixed $\xi$ is exactly proportional to the mass increment $\ud M$ of the halo.

\begin{figure}
    \centering
    \subfigure{
	  \includegraphics[width=1\linewidth, clip]{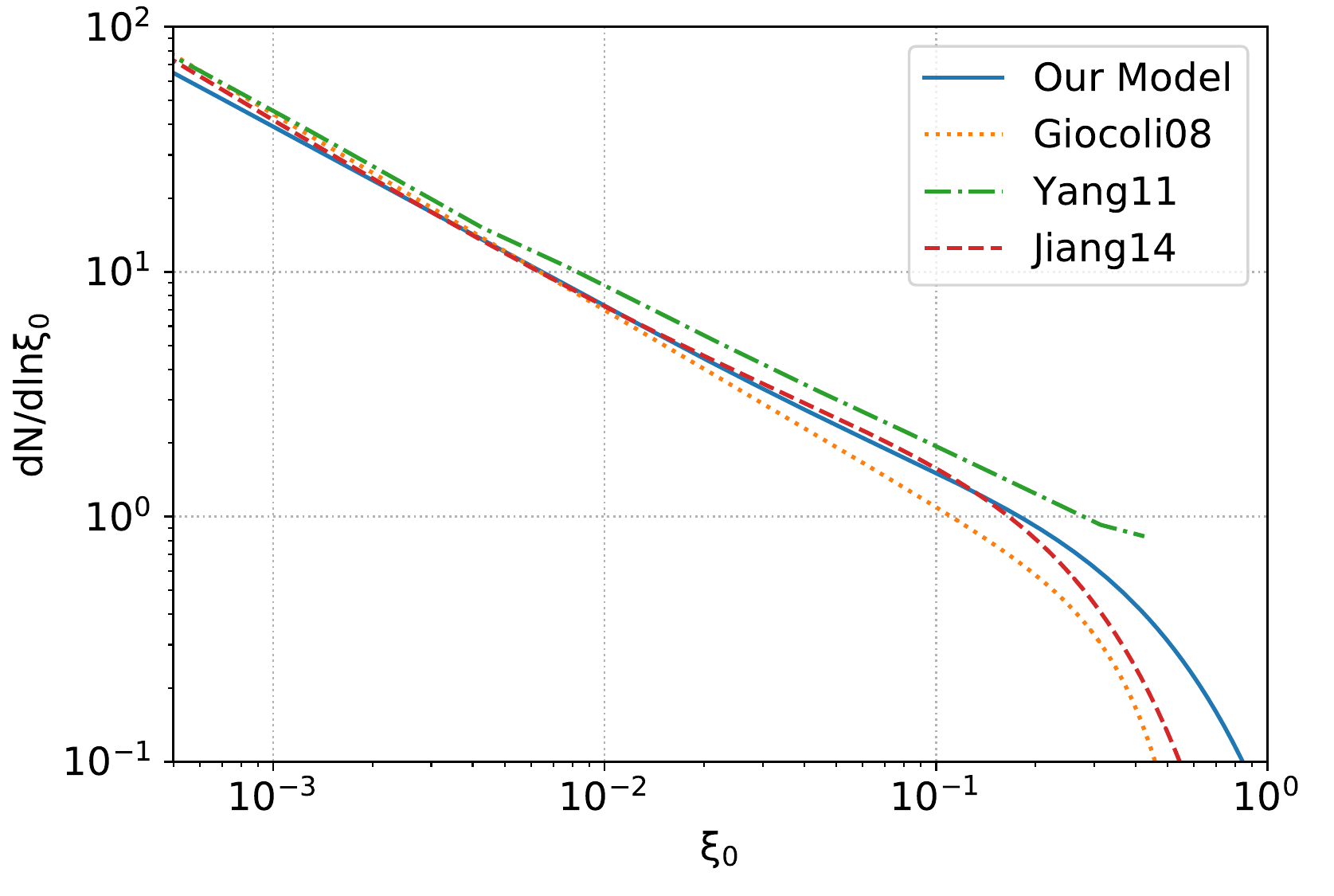}}
   \caption{Un-evolved subhalo mass function. Our model prediction is shown in the blue solid line. The dash-dotted green line, dashed red line and dotted yellow line respectively show the model prediction from \cite{2011ApJ...741...13Y}, the fitting formula of \cite{2014MNRAS.440..193J} and the fitting formula of \cite{2008MNRAS.386.2135G}.}
    \label{fig:uneSMF}
\end{figure}

In Fig.~\ref{fig:uneSMF} we compare our model prediction with some other fitting functions of the USMF. The result of \cite{2008MNRAS.386.2135G} shows an analytical fit of their simulation measurements, for which virial definition is adopted for the halo mass and merger.
The model of \cite{2011ApJ...741...13Y} is a semi-analytical model which evolves progenitor halos from an empirically modified extended PS formula and the mass assembly model of \cite{2009ApJ...707..354Z}. 
The result of \cite{2014MNRAS.440..193J} is a fitting function for the simulation measurements of \cite{2009arXiv0908.0301L}, which also adopts the virial definitions of quantities. 
For the mass ratio larger than $10^{-2}$, our model prediction locates between the results of \cite{2008MNRAS.386.2135G} and \cite{2011ApJ...741...13Y}. Among these three curves, our result shows an overall excellent agreement with \cite{2014MNRAS.440..193J}, but is higher at the large mass end ($\xi_0>0.3$), which indicates the massive subhalos is more sensitive to different treatments of the halo merger than the low mass ones. 
At the very low-mass end $\xi_0<10^{-3}$, our prediction is slightly lower than all three models, which might be due to the resolution effect in our sample.
It is worth re-calibrating our model with simulation results in a larger dynamical range in the future. \citet{2018MNRAS.474..604H} provided universal fittings of the unevolved subhalo mass functions including contributions from all levels of subhalos. However, as our merger rate only accounts for first level subhalos, we do not compare with it here.  

Even though the universality of the USMF has been proposed for a long time, the universality of the merger rate has not been known before. So we believe it is important that we explicitly point out this equivalence both analytically and experimentally, thus unifying our understanding of the two.

\subsection{Estimating the Absolute Merger Rate}
\begin{figure*}
    \centering
    \subfigure{
     \includegraphics[width=0.48\linewidth,clip]{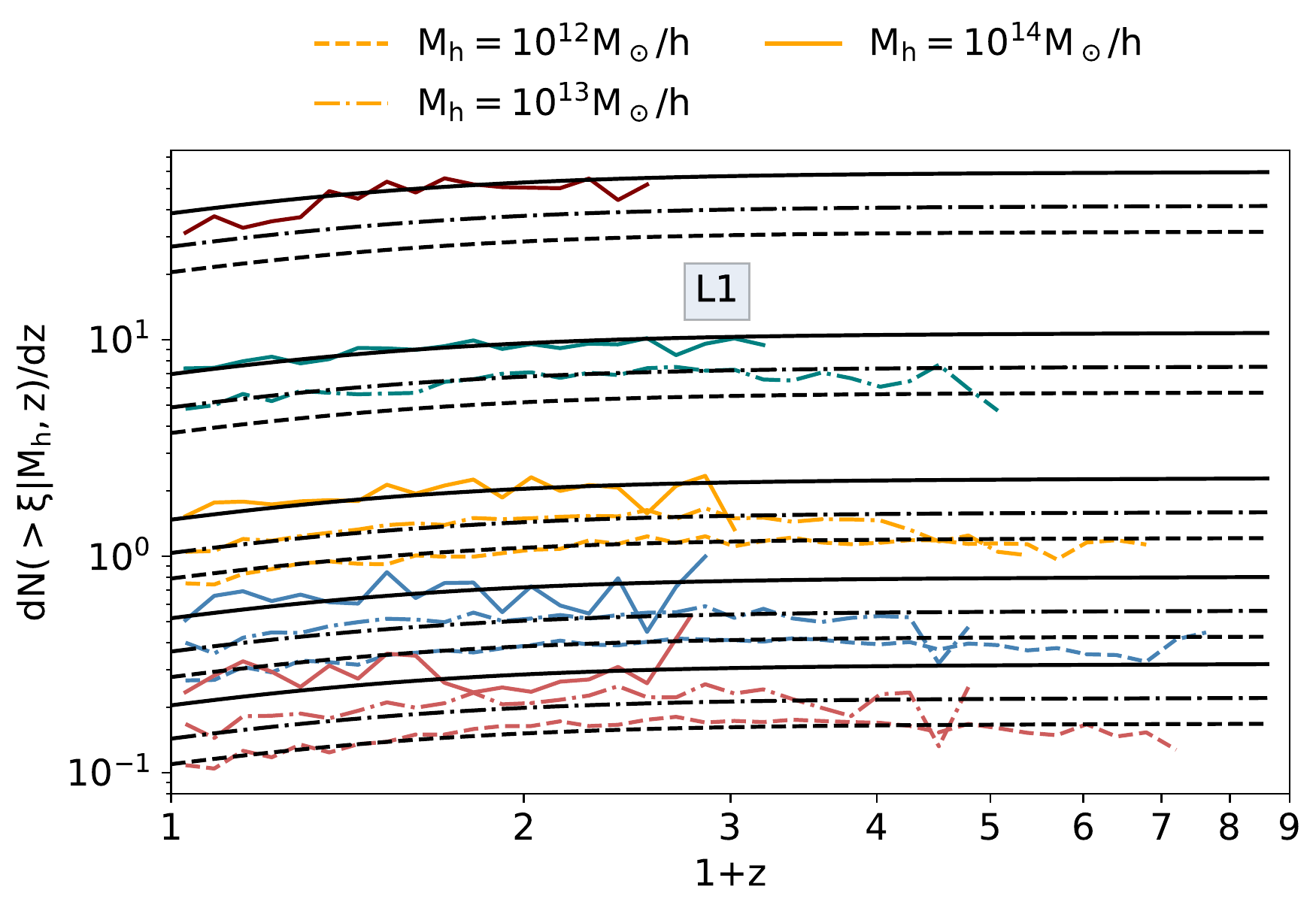}}
    \subfigure{
     \includegraphics[width=0.48\linewidth,clip]{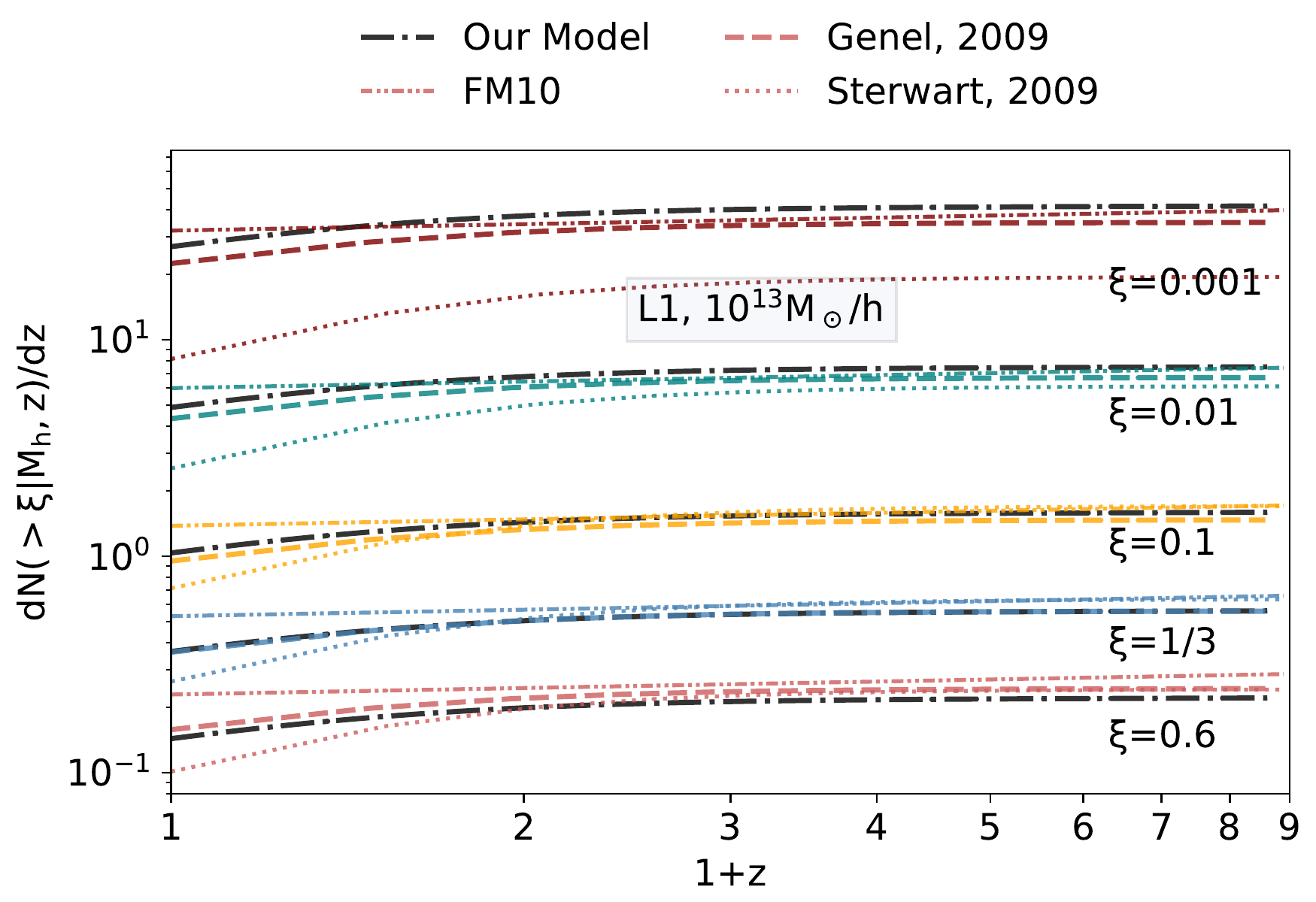}}
     \centering
    \subfigure{
     \includegraphics[width=0.48\linewidth,clip]{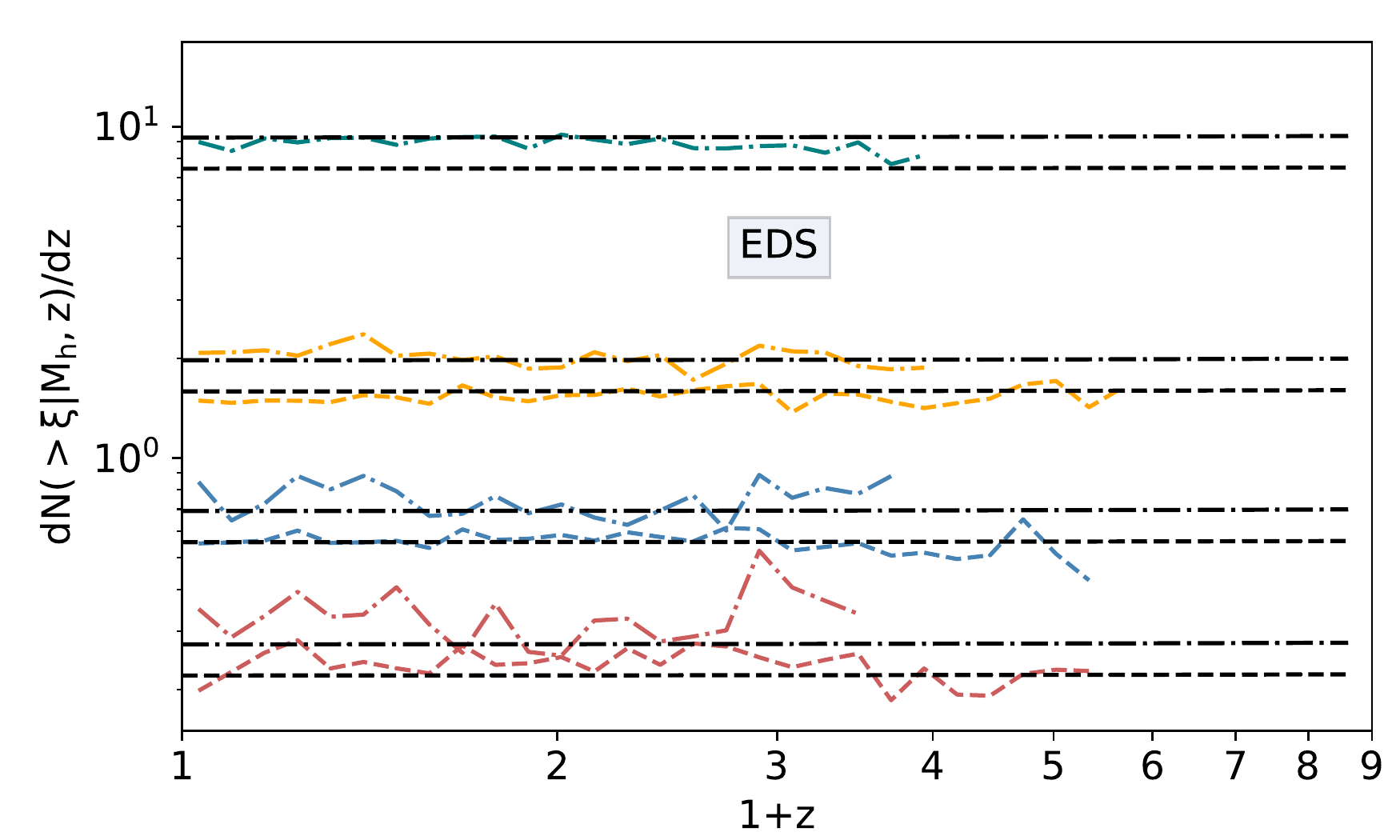}}
    \subfigure{
     \includegraphics[width=0.48\linewidth,clip]{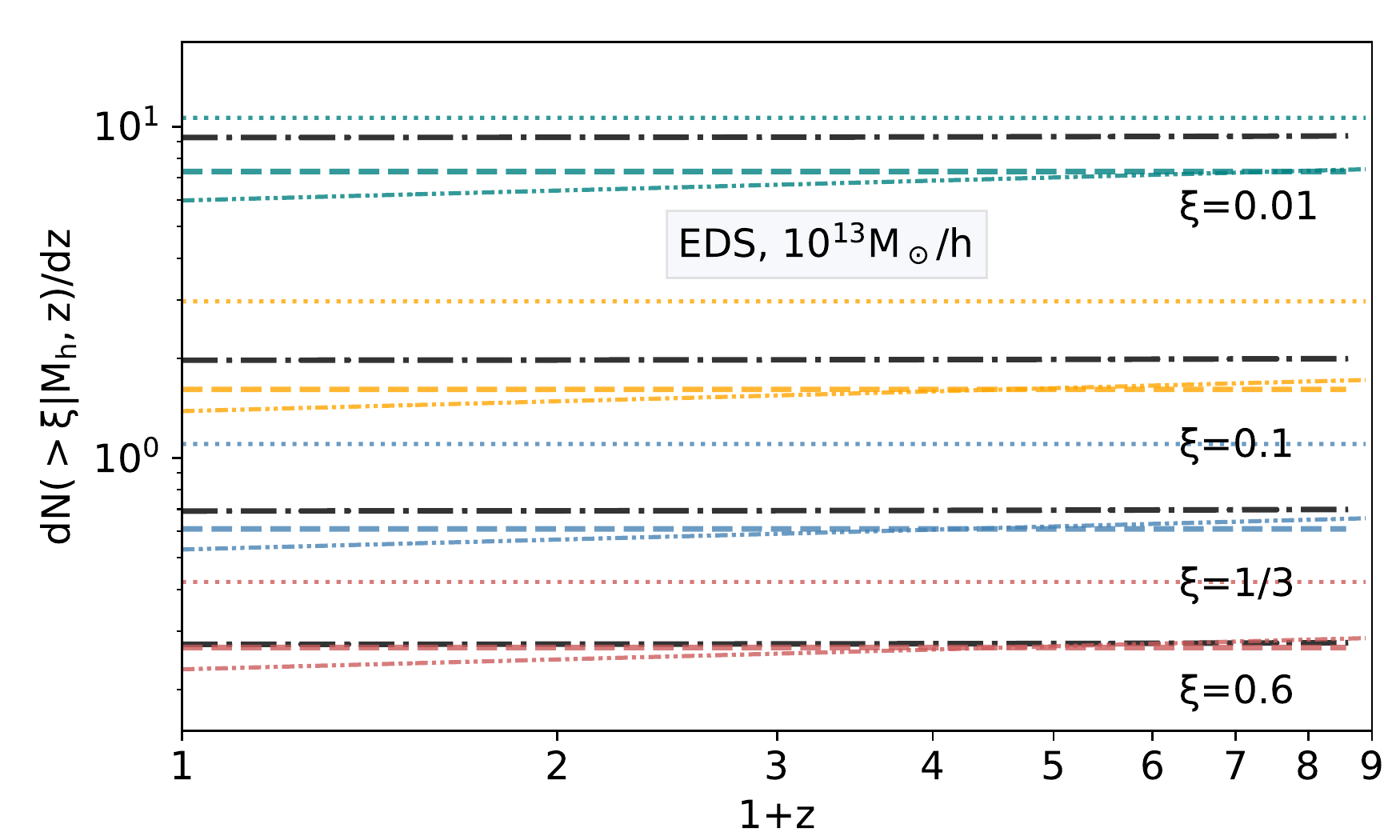}}
   \caption{The halo merger rate per unit redshift $dN_{\mathrm{merge}}(\ge\xi)/dz$. Upper two panels: the halo merger rate in L1 cosmology. Lower two panels: the halo merger rate in EDS cosmology. In the left panels we compare the simulation results with our model predictions, where the latter are shown in black lines. In the right panels we compare our model predictions (black lines) for $M_h=10^{13}M_\sun/h$ with other works, where FM10 is shown in the dashed-dotted line, G09 shown in the dashed line and S09 shown in the dotted line.}
    \label{fig:fm-nm}
\end{figure*}

The specific merger rate shows us that the halo is built up self-similarly, indicating the merger rate per unit redshift depends on redshift only through the halo mass. Thus, one can decompose the  merger rate of a host halo at ($M_{obs},\,z_{obs}$) into two independent terms:
\begin{equation}
\label{eq:dndzobs}
\begin{split}
\frac{\mathrm{d}N_{\mathrm{merge}}(>\xi|M_{obs},z_{obs})}{\mathrm{d}z_{obs}}=
 \frac{\mathrm{d}N_{\mathrm{merge}}}{\mathrm{d}\log M_{obs}}\times\frac{\mathrm{d}\log M_{\mathrm{obs}}}{\mathrm{d}z_{\mathrm{obs}}}.
\end{split}
\end{equation}
The first term $\mathrm{d}N_{\mathrm{merge}}(>\xi|M_{\mathrm{obs}},z_{\mathrm{obs}})/\mathrm{d}\log M_{\mathrm{obs}}$ is a constant over time and can be obtained with EQ.\ref{eq:DS-fit}.
To predict the merger rate, we use a widely adopted model for the MAHs of dark matter halos, for which we refer the readers to \cite{2009ApJ...707..354Z} for more details.
This MAH model is accurate and universal over large dynamical ranges: the same set of model parameters work well for different cosmological models and for halos of different masses at different redshifts.
To get the instantaneous mass grow rate $\mathrm{d}\log M_{\mathrm{obs}}/\mathrm{d}z_{\mathrm{obs}}$  we only need to trace the MAH with one-step backward. We set the shift parameter to zero in the MAH model. 

In Fig.\ref{fig:fm-nm} we show our model predictions of $\mathrm{d}N(> \xi)/\mathrm{d}z$.  In the first panel, we compare our results with the simulation results of L1(\&2).
We remove lines suffering from resolution effect. By adding a multiplicative factor of 0.82, our model of $\mathrm{d}N(> \xi)/\mathrm{d}z$ gives a good description of the simulation results. The multiplicative factor here is mainly due to the difference between our MAH and the model of \cite{2009ApJ...707..354Z}, which may be attributed to our different halo samples, the different ways in getting the average mass growth ratio $\Delta \log M$ \footnote{We measure the average mass growth ( $\Delta \log M=\langle \log (M(z_1)/M(z_2) \rangle$ ) while \cite{2009ApJ...707..354Z} ( $\log  [\mathrm{Median}( M(z_1))/ \mathrm{Median}(M(z_2))]$ ) measured the median MAH.} as well as the ways in computing the halo mass.
In the upper right panel, we compare our model for halos with $M=10^{13}M_\odot/h$ to three other studies. The dashed dotted lines show the fitting formula given by FM10:
\begin{equation}
\label{eq:FM10}
\mathrm{d}N_{merge}/\mathrm{d}z/\mathrm{d}\xi=A\left(\frac{M}{M_\star}\right)^\alpha\xi^\beta \mathrm{exp}\left[(\frac{\xi}{\widehat{{\xi}}})^\gamma\right](1+z)^\eta,
\end{equation}
where $(\alpha,\beta,\gamma,\eta, A,\widehat{\xi})$= (0.133, $-1.995$, 0.263, 0.0993, 0.0104, 9.72e-3), $M_\star=10^{12}M_\odot$.
In addition, we also show comparisons with G09 and S09. G09 is also based on the Millennium simulation for which the difference in their fitting formula from EQ.\ref{eq:FM10} is that $d\delta_c/dz$ has been adopted to describe the redshift dependence. It resulted in a different set of parameters compared to FM08 by introducing a novel merger tree construction: ($\alpha,\beta,\gamma,\eta, A,\widehat{\xi}$)= (0.12, -1.8, 0.5, 1, 0.06, 0.4). 
S09 has adopted a cosmology of $\Omega_m=0.3$ and $\sigma_8=0.9$, and a slightly different fitting formula 
$\mathrm{d}N/\mathrm{d}z(>\xi)=0.27(\mathrm{d}\delta_c/\mathrm{d}z)^2M_{12}^{0.15}\xi^{-0.5}(1-\xi)^{1.3}$, with  $M_{12}$ being the mass in units of $10^{12}M_\sun/h$. These two results are respectively shown in dashed lines and dotted lines in the figure. To make a fair comparison, we set the cosmologies of these models the same as L1. On the whole, our model is broadly consistent with all of these three works. In more details, our model shows the best consistency with G09 in both the shape and amplitude. Compared with FM10, our results are consistent with theirs at $z>1$, but show a more rapid increase with redshift at $z<1$. The S09 results show a lower merger rate for $\xi\le0.01$ and a more rapid increase with redshift for $z<1$ than other models.
Overall, the consistency between our results and these models confirms the accuracy of our model of $\mathrm{d}N(\xi)/dz$. 

As a further test, we show the results of the EDS cosmology in the lower panels. From the third panel we can find that our model is still able to describe the simulation results very well, for which the same multiplicative factor of 0.82 is adopted. While all the other formulas in the fourth panel show obvious discrepancies in amplitude from our model.

To sum up, instead of providing a fitting formula to the simulation results of the  merger rate $\mathrm{d}N(\xi)/dz$, we propose a model by combining the specific merger rate with the MAH model of \cite{2009ApJ...707..354Z}, which is more clear in physics. More importantly, our model is universal and enables us to estimate the merger rate in different cosmologies.

\subsection{Estimating the History of Major Mergers}
\label{subsec:major}
\begin{figure}
    \centering
    \subfigure{
	  \includegraphics[width=1\linewidth, clip]{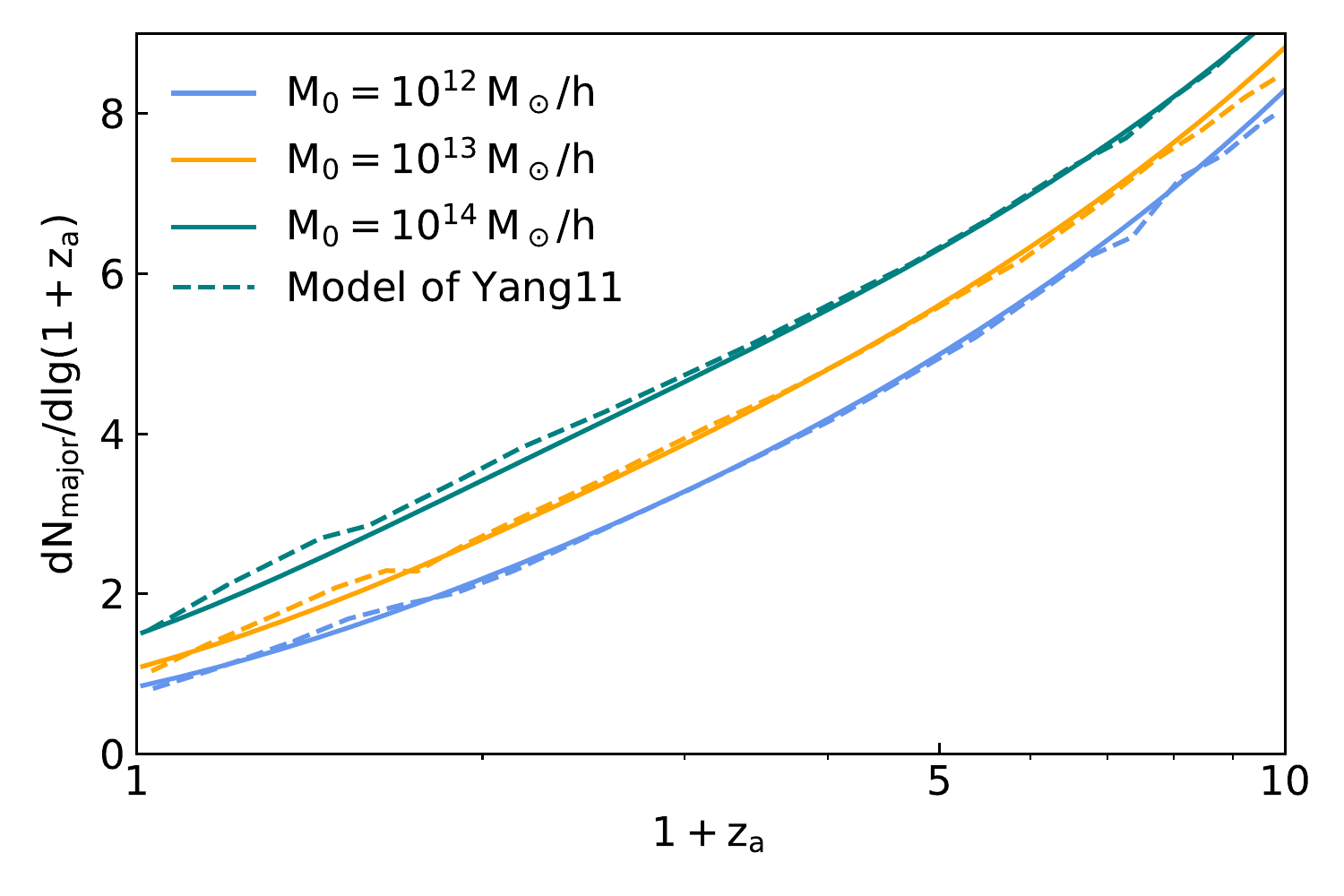}}
   \caption{Historical major merger events per unit $\log (1+z_a)$ as a function of accretion time $z_a$ of present-day halos. Our estimations are shown in the solid lines, while the model predictions of \cite{2011ApJ...741...13Y} are shown in the dashed line. The comparison is done for halos with $M_0=$ $10^{12}M_\odot/h$, $10^{13}M_\odot/h$ and $10^{14}M_\odot/h$.  }
    \label{fig:cmp-yang}
\end{figure}

By applying Equation~\eqref{eq:dndzobs} along the MAH of a given halo, one can immediately predict its merger rate history. For a halo with mass $M_0$ and $z_0$, its MAH can be specified with the \citep{2009ApJ...707..354Z} model, so that its merger rate history is given by 
\begin{multline}
\mathrm{d}N_a(\xi,z_a|M_0,z_0)/\mathrm{d}z_a=\frac{\mathrm{d}N(\xi,z_a|M_0,z_0)}{\mathrm{d}\log M_a(z_a| M_0,z_0)}\times\frac{\mathrm{d}\log M_a}{dz_a},
\end{multline}
where $z_a$ the accretion time, and $M_a$ the main branch mass at $z_a$. According to our conclusion, $\mathrm{d}N(\xi,z_a|M_0,z_0)/\mathrm{d}\log M_a$ is a constant over redshift for a given $\xi$.

In Fig.\ref{fig:cmp-yang} we compare our model predictions of the major merger rates ($\xi\ge1/3$) with the analytical model for the accretion of subhalos developed in \cite{2011ApJ...741...13Y} (here after Y11), by adopting the same MAH model of $M_a(z_a,M_0)$ \citep{2009ApJ...707..354Z}. Here we still set the shift parameter to zero in the model. We find that by considering a multiplicative factor 0.89, we can get a consistent result with  Y11. The difference in amplitude may be due to our different treatments in measuring the halo mass and merger rate.

\section{Comparison with the EPS Prediction}\label{sec:EPS}
\begin{figure}
    \centering
    \subfigure{
	  \includegraphics[width=1\linewidth, clip]{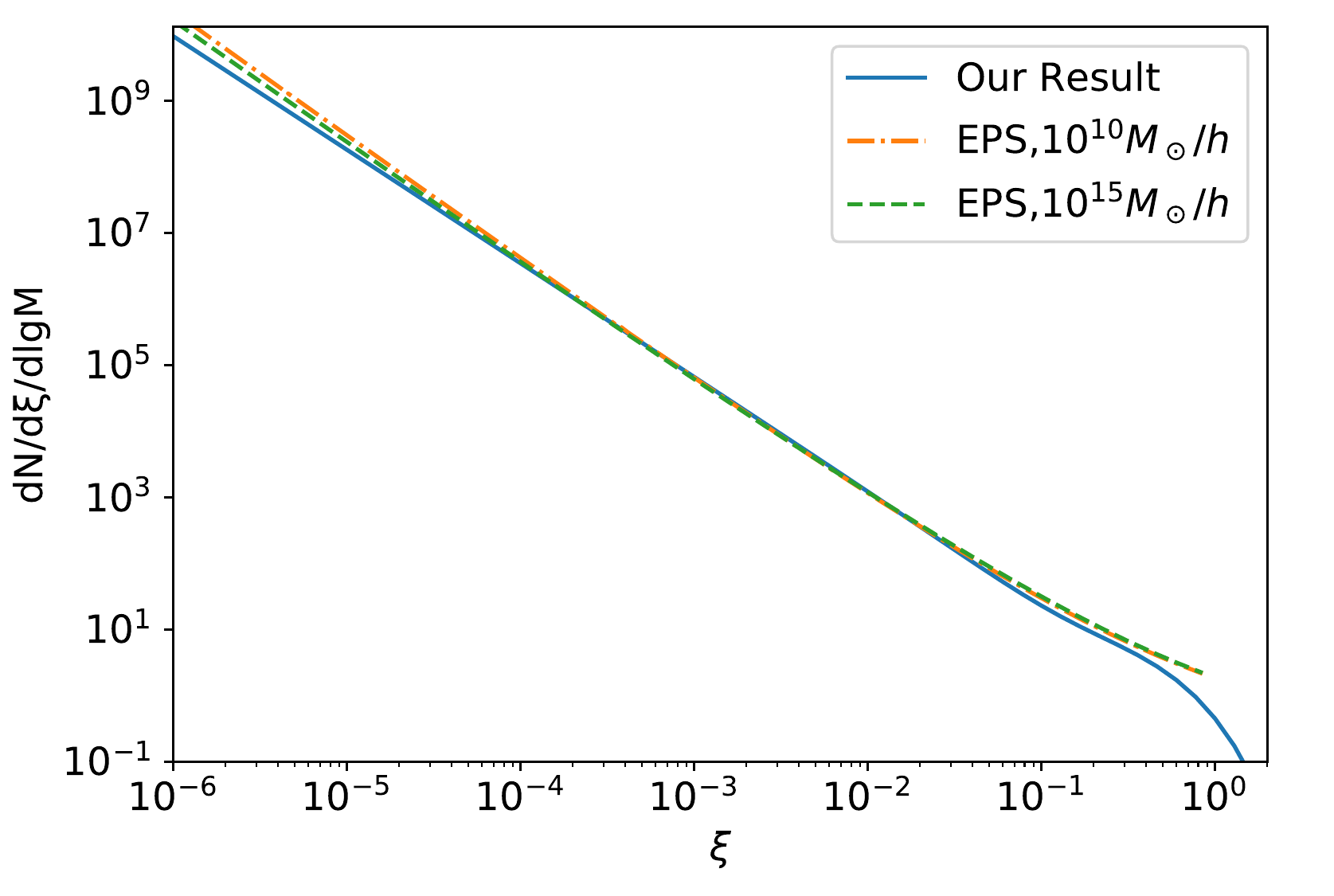}}
   \caption{The specific merger rate estimated by the EPS framework. The orange and green line show the merger rates as a function of the progenitor merger ratio for two descendat halo masses: $10^{10}$ and $10^{15}M_\odot/h$. The blue dashed line shows the simulation result.}
    \label{fig:cmp-eps}
\end{figure}

In this section, we compare our results with predictions from the EPS formalism. 
Denoting $M_0$ as the descendant mass of a halo at $z_0$, $M_1$ the mass of its progenitor at $z_1>z_0$ ($M_1<M_0$), the number weighted conditional mass function of $M_1$ is given as \citep{1993MNRAS.262..627L,2000MNRAS.319..168C}:
\begin{equation}
\label{eq:app1}
    \phi(M_1,z_1|M_0,z_0)=-\frac{M_0}{M_1}\frac{\Delta \omega}{\sqrt{2\pi}\Delta S^{3/2}}\mathrm{exp}[-\frac{(\Delta \omega)^2}{2\Delta S}]\frac{dS_1}{dM_1},
\end{equation}
where $\Delta w=w(z_1)-w(z_0)$, $\Delta S=S(M_1)-S(M_0)$, $w(z_1)$ the critical over-density at $z_1$. For binary mergers, the descendant mass $M_0$ can be written as the sum of two progenitors  $M_0=M_1+M_2$. Consequently, the average merger rate per descendant halo can be related to the conditional mass function as \citep{10.1093/mnras/289.1.66,2008MNRAS.389.1521Z,2008MNRAS.387L..13Z} 
\begin{equation}
    \frac{\Delta N(\xi)}{\Delta z\Delta \xi}=\phi(M_i,z_1|M_0,z_0)\frac{M_0}{(1+\xi)^2\Delta z},
\end{equation}
where $M_i$ can be either of $M_1$ or $M_2$, $\Delta z=z_1-z_0$ is the time step and $\xi=M_2/M_1$ is the progenitor mass ratio. Due to the asymmetry of EQ.\ref{eq:app1} of the EPS model\citep{2008MNRAS.389.1521Z}, here we assign $M_i$ as the less massive progenitor $M_2$. Meanwhile, the mass change of the host $\Delta \log M$ between $z_1$ and $z_0$ can be obtained as $\log (1+\sum_j\xi_j\Delta N(\xi_j))$, where $\xi_j\Delta N(\xi_j)$ is the relative mass change of the host halo, $\Delta M(\xi_j)/M_1$, induced by mergers of $\xi_j$. For a small time step ($\Delta z\to0$), $\Delta \log M$ is proportional to $(\sum_j\xi_j\Delta N(\xi_j))^{-1}$.

In Fig.\ref{fig:cmp-eps}, we show the EPS estimations of $f(\xi,M_0)$ for descendant halos at $z_0=0$ with $\Delta z=0.02$.
We can find that the specific merger rate estimated for $M_0=10^{10}M_\odot/h$ is almost the same as that for $M_0=10^{15}M_\odot/h$. Although slightly more minor mergers and fewer major mergers are found for the former, the difference is tiny at $\xi>10^{-4}$. For a small time step, the exponential term in the conditional function diminishes. 
The mass related term $\mathrm{d}S_i/\mathrm{d}M_i/\Delta S^{3/2}$ is found to have a rough power law relation with the progenitor mass $M_i$, which thus helps to remove the dependence of $f(\xi)$ on the descendant mass $M_0$ and leads to a quite universal form. The same conclusion holds for $z_0>0$.

It is encouraging to see that our simulation result is consistent with the EPS predictions for $10^{-4}<\xi<10^{-1}$. While the EPS results are derived numerically without a simple expression, our empirical law from the simulation provide a much more direct and convenient way for predicting the merger rate. As our result summarizes the simulation concisely, it can also serve as an important benchmark for calibrating theoretical models of halo growth. In Section~\ref{sec:summay} we will also briefly discuss the connections of our result to alternative theoretical models such as the CUSP formalism~\citep{2021ApJ...914..141S}.




\section{Conclusion \& Discussion}
\label{sec:summay}
In this work, we study the merger rate of dark matter halos over time, mass and cosmology using a set of $N$-body simulations. Unlike previous studies, we focus on the specific merger rate which we find to be universal. We propose a fitting formula to the specific merger rate, and compare our model with other statistics on halo mergers as well as with theoretical predictions. In particular, we have shown that our universal specific merger rate is equivalent to a universal unevolved subhalo mass function (USMF). 
Our main results can be summarized as follows:

\begin{itemize}
\item We define the specific merger rate by normalizing the instantaneous merger rate with the logrithmic mass growth rate of the host. 
We find that this specific merger rate is universal for different halo mass, redshift and cosmology.
The universality of the specific merger rate reveals a strong self-similarity in the merger-driven growth of halos, such that the merger rate scales with the mass growth rate, with the same mass ratio distribution in the progenitors at each step. 
\item Consequently, the absolute merger rate $\mathrm{d}N(\xi|z,M)/\mathrm{d}\xi/\mathrm{d}z$ depends on redshift only through the halo mass variable, which can be easily predicted from the specific merger rate combined with the universal MAH model of \cite{2009ApJ...707..354Z}. 
\item The same conclusion holds for present-day halos through their mass accretion histories. 
\item The universal specific merger rate naturally results in a universal USMF that only depends on the mass ratio but not on the host halo mass, redshift or MAH. This is a direct consequence of our finding that the halo accretion closely traces the mass accretion, resulting in a final progenitor distribution that is independent on the path of the mass growth, in line with the ``unbiased accretion" picture discussed in \citet{2016MNRAS.457.1208H}.
\end{itemize}

The above conclusions are valid at least for halos over the range of mass  $[10^{12}, 10^{14}]\,M_\odot/h$ and merger ratio $\xi>0.01$. Compared with previous works, the specific merger rate defined in this work is simpler in the description and much easier to interpret physically, with no dependence on the cosmology. In particular, our formalism has the appealing advantage that it is mathematically equivalent to the universal USMF. As a result, our findings on the universality of the merger rate also substantially extend our understanding of the universality of the USMF over time and cosmology.

Together with the universal USMF, we expect our findings can find wide applications in many problems. As our specific merger rate describes the simulation result in a concise and convenient way, it can serve as an important benchmark for developing and calibrating theoretical models of halo growth. For example, we have demonstrated that the EPS model can produce consistent results with ours, but in a much more involved way with subtle differences.
Our model can also be used in conjugation with the universal distribution of infall orbits\citep{2020ApJ...905..177L} to generate initial conditions for halo mergers in semi-analytical or numerical studies of halo and galaxy formation. 

In this work we only focus on the mass distribution of mergers without discussing the spatial distribution of the merger remnants, i.e., subhalos. Despite this, our findings may be used to provide insights on the universal spatial distribution of unevolved subhalos~\citep{2016MNRAS.457.1208H} as well. In fact, if approximating the progenitor distribution (Equation~\eqref{eq:DS-fit}) at the low mass end with a single power law of $\sim a\xi^{-2}$, we can get $\mathrm{d}N/\mathrm{d}m/\mathrm{d}M\sim a m^{-2}$. In this case, the number of subhalos accreted is exactly proportional to the mass accreted irrespective of accretion time, so that the accretion of subhalos is ``unbiased" relative to mass accretion. This is exactly the ``unbiased" accretion picture described in \citet{2016MNRAS.457.1208H}, which would result in the distribution of subhaloes tracing dark matter distribution dynamically. 
Alternatively, if we assume halos are built-up essentially shell by shell from inside-out as in the CUSP formalism \citep{2021ApJ...914..141S,SS21},\footnote{Note the CUSP formalism itself does not rely on the assumption of inside-out growth, and can be applied to both purely accreting halos and systems experiencing major mergers.} the number of mergers over time can be translated to the spatial distribution of accreted subhalos over radius.
Under the same approximation of a $\sim \xi^{-2}$ distribution at the low mass end, the derived spatial distribution can become separable from the mass distribution, reproducing the \citet{2016MNRAS.457.1208H} model. We note that the CUSP formalism can also derive these distributions completely independently. However, similar to the EPS model which we discussed above, the derived distributions may have certain deviations from our empirical results. Our results thus could be potentially used to test and calibrate such theoretical models.


Finally, our work can be further improved in several aspects.
For example, to test our conclusion for halos over a larger dynamical range, we need simulations with larger box size and better resolution in the next step.   Besides, the infall rate and splashout rate themselves are also important quantities that can be studied separately.
In fact, we find that although the normalized infall rate decline with redshift, it has no dependence on the halo mass. And the same trend is found for the splashout rate.
But the measurement of splash rates are very easily influenced by the simulation resolution, as the masses of splashed halos are usually low, and thus introduce resolution effect to our statistics. It also remains interesting to generalize our analysis to study the galaxy merger rate which are found to be qualitatively similar\citep{2015MNRAS.449...49R}. Last but not the least, it could be interesting to study the merger rate under alternative physical definitions of the halo boundary (e.g., depletion radius, \citealp{FH21}) which may be able to provide more consistent and natural definitions of the ``merger" among haloes. We hope to study these issues in more details in the future.

\section{ACKNOWLEDGEMENTS}
We thank for useful discussions with Youcai Zhang.
This work is supported by NSFC (11973032, 11890691, 11621303), National Key Basic Research and Development Program of China (No.2018YFA0404504), 111 project No. B20019, and the China Manned Space Project with NO.  CMS-CSST-2021-A03. FYD is supported by a KIAS Individual Grant PG079001 at Korea Institute for Advanced Study.



\bibliography{merger}
\end{document}